\documentclass[%
reprint,
superscriptaddress,
showpacs,preprintnumbers,
amsmath,amssymb,
aps,
]{revtex4-1}

\usepackage{color}
\usepackage{graphicx}
\usepackage{dcolumn}
\usepackage{bm}

\begin {document}
\preprint{APS/123-QED}

\title{Transition from distributional to ergodic behavior in an inhomogeneous diffusion process: 
Method revealing an unknown surface diffusivity}

\author{Takuma Akimoto}
\email{akimoto@z8.keio.jp}
\affiliation{%
  Department of Mechanical Engineering, Keio University, Yokohama, 223-8522, Japan
}%

\author{Kazuhiko Seki}
\affiliation{%
Nanosystem Research Institute, AIST, Tsukuba, 305-8565, Japan
}%


\date{\today}

\begin{abstract}
Diffusion of molecules in cells plays an important role in providing a biological reaction on the surface 
by finding a target on the membrane surface. The water retardation (slow diffusion) near the target
assists the searching molecules to recognize the target.  
Here, we consider effects of the surface on the diffusivity in three-dimensional diffusion processes, where diffusion on the surface 
is slower than that in bulk. We show that the ensemble-averaged mean square displacements increase linearly with 
time when the desorption rate from the surface is finite even when the diffusion on the surface is subdiffusion. Moreover, 
this slow diffusion on the surface affects the fluctuations of the time-averaged 
mean square displacements (TAMSDs).  We find that fluctuations of the TAMSDs remain large when the measurement time 
is smaller than a characteristic relaxation time, and decays according to an increase of the measurement 
time for a relatively large measurement time.   Therefore, we find a transition from non-ergodic (distributional) to ergodic 
diffusivity in a target search process. 
Moreover, this fluctuation analysis provides a method to estimate an unknown surface diffusivity. 
\end{abstract}

\maketitle


\section{Introduction} 
Stochastic searching for an unknown position of a target plays an important role in many physical, chemical, and 
biological phenomena.  It has been known that intermittent target search strategies, where there are combinations 
of two different diffusivities (slow and fast diffusivities), is an optimal strategy to find a randomly located object \cite{Benichou2011}.  
In particular, proteins search a target sequence on the DNA using a combination of 3D diffusion and 1D diffusion (sliding on the DNA). 
Many theoretical studies conclude that this protein-DNA search process is facilitated by the 1D diffusion \cite{Coppey2004,Mirny2009,Benichou2011}. 

A combination of slow and fast diffusivities has been observed in many biological phenomena. In enzyme activities, a substrate searches 
a target on the enzyme surface, where a retardation of a diffusivity around the target assists a binding to the target \cite{Grossman2011}. 
Furthermore, it has been found that diffusion of water molecules on the membrane surface exhibits subdiffusion and the origin 
of subdiffusion is a power-law trapping times (continuous-time random walk) and anti-persistence (fractional Brownian motion)
 by molecular dynamics simulations \cite{Yamamoto2013,Yamamoto2014}. Such a slow motion of water molecules plays an 
 important role in enhancing biological reactions on the membrane surface \cite{Ball2011}. 
 Therefore, it is essential to consider a slow motion near targets in efficient target search processes. 
 
 In such a diffusion process, it is interesting to investigate fluctuations of diffusivities in the system because the diffusivity 
 is heterogeneous, indicating that the instantaneous diffusivity fluctuates randomly over time. In fact, diffusivities obtained from 
 single particle trajectories show large fluctuations in heterogeneous diffusion processes such as a diffusion process with 
 space-and time-dependent diffusivity \cite{Fulinski2011,Cherstvy2013,Cherstvy2013a,Jeon2014,Uneyama2014}. Therefore, 
 heterogeneous diffusivities can provide a possible explanation for large fluctuating diffusivities  
 observed in biological transports such as in living cells and on cell membranes \cite{Golding2006,Weigel2011,Jeon2011,Tabei2013}.  
 
 Another explanation of fluctuating diffusivity can 
 be provided by a trap model such as continuous-time random walk (CTRW) \cite{Lubelski2008,He2008,Miyaguchi2013,Metzler2014} and random walk with static disorder 
 \cite{Miyaguchi2011}. In CTRWs, time-averaged mean square displacements (TAMSDs) increase linearly with time even when the waiting-time distribution 
 does not have a finite mean \cite{Miyaguchi2013}. However, when the waiting-time distribution follows a power-law distribution with a divergent mean, 
 the diffusion coefficients 
 remain random even when the measurement time goes to infinity, where TAMSD is calculated by a single trajectory,
 \begin{equation}
\overline{\delta^2(\Delta;t)}\equiv \frac{1}{t-\Delta} \int_0^{t-\Delta} dt' (\bm{r}_{t'+\Delta} -\bm{r}_{t'})^2,
\end{equation}
where $\bm{r}_t$ is a position at time $t$.
  This intrinsic random behavior is characterized by the relative standard deviation 
 (RSD) of the TAMSDs, defined by
 \begin{equation}
\sigma_D(t) =\frac{\sqrt{\langle D_t^2 \rangle - \langle D_t \rangle^2}}{\langle D_t \rangle},
\label{rsd}
\end{equation}
where $\langle \cdot \rangle$ represents an average with respect to realizations, 
$t$ is a measurement time and $D_t$ is the diffusion coefficient, defined as $D_t \equiv \overline{\delta^2(\Delta;t)}/(2d\Delta)$, 
where $d$ is the dimension. When the waiting-time distribution in CTRW follows a power-law distribution with a divergent mean, 
the RSD converges to a non-zero constant even when the measurement time goes to infinity, {\it i.e.}, $t\to \infty$. 
When the waiting-time distribution in CTRW has 
a power-law with an exponential cutoff, the RSD shows a transition from non-ergodic (distributional) behavior to 
ergodic behavior such as $\sigma_D(t) \propto t^{-0.5}$ \cite{Miyaguchi2011a}.

In this paper, we investigate the effective diffusivity and the fluctuations of TAMSDs in three-dimensional random walk with 
a sticky surface, which mimics a diffusion process in cell. We will provide a transition from non-ergodic to ergodic fluctuations 
in TAMSDs, which is similar to CTRW with a power-law distribution with cutoff. These results provide a useful method to 
estimate the diffusivity on the surface.

\section{Model}

We consider a CTRW on 3D lattice as a model of diffusion in cell.
In CTRW, a random walker must wait for a random continuous time to jump. In the model, there are two walls at 
$z=0$ and $z=L$, which represent the membrane surfaces. We consider inhomogeneity on the membrane surfaces (walls), where diffusivity on the 
membrane surfaces is different from that in the bulk. 
Such inhomogeneous diffusion processes are observed in experiments such as diffusion of proteins on the membrane surface
 \cite{Knight2009, Knight2010,Rozovsky2012} and diffusion near a solid-liquid interface \cite{Skaug2014}.
Moreover, the mean first passage time in an inhomogeneous diffusion in a spherical domain has been analytically 
studied \cite{Benichou2010,Benichou2011a,Rupprecht2012,Rupprecht2012a}.

In the bulk, the probability density function (PDF) of the waiting times follows the exponential density 
$\psi_b(\tau) = \tau_b^{-1} \exp(-\tau/\tau_b)$. 
For example, after random waiting times, the random walker can jump to the $x$, $y$, or $z$ direction of $\pm 1$ in equal probability. 
On the membrane surfaces, we consider two different PDFs for waiting times, {\it i.e.}, the exponential and power-law densities, for 
lateral direction ($xy$ plane). That is, the PDFs of the waiting times in the absence of desorption are given by 
$\psi_s(\tau) = \tau_s^{-1} \exp(-\tau/\tau_s)$ and $\psi_s(\tau) 
\propto \tau^{-1-\alpha}$ $(\tau \to \infty)$ for the exponential and power-law densities, 
respectively. For the $z$ direction on the membrane surface, the waiting time is independent of that of $xy$ plane and follows 
the exponential density, {\it i.e.}, $\phi(\tau) = \tau_z^{-1} \exp(-\tau/\tau_z)$. On the membrane surfaces, the random walker jumps to $x$ or $y$ 
direction with equal probability 
 if waiting time for $xy$ plane is smaller than that for the $z$ direction, and it can desorb from the membrane surface 
otherwise.  

\begin{figure}
\includegraphics[width=.9\linewidth, angle=0]{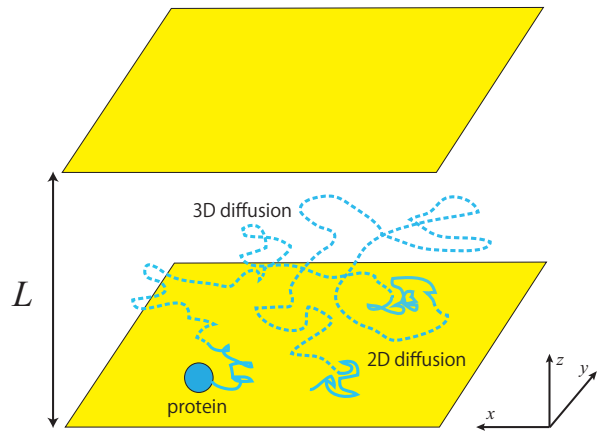}
\caption{Schematic figure of our model. Diffusivity on the surfaces is different from that in bulk.}
\end{figure}

\section{Mean square displacement}

Here, we analytically calculate the mean square displacement (MSD) for the cases of exponential and power-law waiting time distributions 
on the membrane surface.  We define the ratio between total residence time on the membrane surfaces and total measurement time as $T_s$:
\begin{equation}
T_s \equiv \lim_{t\to\infty} \frac{t_s^1 + \cdots + t_s^{N_t}}{t},
\end{equation}
where $t_s^i$ is the $i$th residence time on the membrane surfaces and $N_t$ is the number of visits to the membrane surfaces until time $t$. 
We note that the ratio is given by the mean residence time on the membrane surfaces $\mu_s$ and the mean return time to the membrane
surfaces after desorbing 
from the walls $\mu_b$: $T_s =\mu_s/(\mu_s + \mu_b)$. Furthermore, $\mu_b$ is given by $\mu_b = 3(L-1)\tau_b$ because 
the mean waiting time for the $z$ direction in the bulk is given by $3\tau_b$ and the mean first passage time to 
$z=0$ or $z=L$ starting from $z=1$ or $z=L-1$ is given by $L-1$ times the mean waiting time for an usual random walk in one dimension. 
As shown in Appendix~A, the mean residence time on the membrane surfaces is given by $\mu_s = \tau_z$ both in the cases of the exponential and power-law 
 waiting-time PDF on the $xy$ plane. More precisely, the PDF of the residence times on the membrane surface 
is the exactly same form as $\phi(\tau)$. 

The MSD for the $x$ direction is given by
\begin{equation}
\langle x_t^2 \rangle =  \langle ( \Delta x_{t_s^1} + \Delta x_{t_b^1} + \cdots + \Delta x_{t_s^{N_t}} + \Delta x_{t_b^{N_t}} + \Delta_t)^2 \rangle
\end{equation}
where $\Delta x_{t_s^i}$ is the displacement for $x$ direction during the $i$th residence on the membrane surfaces,
$\Delta x_{t_b^i}$ is the displacement for $x$ direction during the $i$th residence in the bulk, and $\Delta_t$ is a correction term associated with 
the last step, where 
$\langle \Delta_t^2 \rangle$ 
can be negligibly small when $t$ is sufficiently large, {\it i.e.}, $\langle \Delta_t^2 \rangle/ \langle x_t^2 \rangle \to 0$ 
as $t\to \infty$.  
The displacements are given by 
$\Delta x_{t_s^i} = x_{t_i + t_s^i} - x_{t_i}$ and $\Delta x_{t_b^i} = x_{t_{i+1}} - x_{t_{i+1} - t_b^i}$ if the random walker is initially located 
on the membrane surface, where $t_i=\sum_{k=1}^{i-1} (t_s^k + t_b^k)$. 
Because the displacements are independent of each other, we have
\begin{equation}
\langle \Delta x_{t_s^i}\cdot \Delta x_{t_s^j} \rangle = \langle \Delta x_{t_b^i}\cdot \Delta x_{t_b^j} \rangle = 0
\quad (i\ne j)
\end{equation}
and 
\begin{equation}
\langle \Delta x_{t_s^i}\cdot \Delta x_{t_b^j} \rangle =0.
\end{equation}
It follows that the MSD is given by
\begin{equation}
\langle x_t^2 \rangle \sim  \sum_{n=1}^{\infty} \Pr (N_t =n) \sum_{i=1}^n \{ \langle \Delta x_{t_s^i}^2 \rangle + 
\langle \Delta x_{t_b^i}^2 \rangle \}
\label{msd}
\end{equation}
for $t\to\infty$. 

We note that an equilibrium probability with respect to the $z$ direction exists
because there is a confinement for the $z$ direction and the mean residence time on the membrane surface is finite. 
In particular, the equilibrium probability is given by 
\begin{equation}
\rho_{eq}(z) = \left\{
\begin{array}{ll}
\dfrac{\mu_s}{2(\mu_s + \mu_b)} \quad &(z= 0)\\
\\
\dfrac{\mu_b}{(L-1)(\mu_s + \mu_b)} \quad &(0<z<L)\\
\\
\dfrac{\mu_s}{2(\mu_s + \mu_b)} \quad &(z= L).\\
\end{array}
\right.
\end{equation}

\subsection{Exponential waiting-time distribution}

The diffusion coefficients, defined by $D=\langle {\bm r}(t)^2 \rangle/(2dt)$,
 in the bulk ($d=3$) and on the membrane surfaces ($d=2$) are given by $1/(6\tau_b)$ and $1/(4\tau_s)$, respectively. 
We use the following  notation: $D_b=1/(6\tau_b)$ and $D_s=1/(4\tau_s)$. 
Because the initial condition for the $z$ position is in equilibrium, 
we have $\langle N_t \rangle = t/(\mu_s + \mu_b)$ \cite{Cox}. Using $\langle \Delta x_{t_b^i}^2 \rangle=2D_b t_b^i$
and $\langle \Delta x_{t_s^i}^2 \rangle=2D_s t_s^i$, we obtain the MSD:  
\begin{equation}
\langle x_t^2 \rangle \sim  2\left( \frac{D_s \mu_s}{\mu_s + \mu_b} + \frac{D_b \mu_b}{\mu_s + \mu_b} \right)t
\end{equation}
for $t\to \infty$.
Because 
the MSD for the $y$ direction is the same as that for the $x$ direction, 
the lateral MSD (the MSD on the $xy$ plane) is given by
\begin{equation}
\langle x_t^2 + y_t^2\rangle \sim  4\left( \frac{D_s \mu_s}{\mu_s + \mu_b} + \frac{D_b \mu_b}{\mu_s + \mu_b} \right)t
\label{msd_exp}
\end{equation}
for $t\to \infty$.  Therefore, the effective lateral diffusion coefficient, 
$D_{\rm eff}\equiv \langle x_{t}^2 + y_{t}^2\rangle/(4t)$ as $t\to \infty$, is given by
\begin{equation}
D_{\rm eff} =\frac{D_s\mu_s + D_b \mu_b}{\mu_s + \mu_b}.
\end{equation}
The result is consistent with  that for diffusion in multilayer media \cite{Berezhkovskii2006}.

\subsection{Power-law waiting time distribution}
Equation~(\ref{msd}) can be used even when the waiting time distribution is not exponential. 
To investigate an effect of anomalous diffusion (subdiffusion), {\it i.e.}, the MSD grows sublinearly with time, on the membrane surface, 
we consider the following power-law waiting-time distribution on the surfaces, $\psi_s(\tau) \sim \frac{\alpha}{\tau_s} \left(\frac{\tau}{\tau_s}\right)^{-1-\alpha}$ 
$(\tau\to\infty)$ with $\alpha \leq 1$.
In this case, the MSD on the membrane surface shows subdiffusion:
\begin{equation}
\langle x_{t_s}^2 + y_{t_s}^2\rangle \cong D'_s t_s^\alpha,
\end{equation}
where $D'_s = 1/\Gamma(1-\alpha)\Gamma(1+\alpha)\tau_{s}^\alpha$. As in the calculation of the exponential 
case, we have 
\begin{equation}
\langle x_t^2 + y_t^2\rangle \sim  \left( \frac{D'_s \langle \tau_s^\alpha \rangle}{\mu_s + \mu_b} + \frac{2D_b \mu_b}{\mu_s + \mu_b} \right)t,
\label{msd_power}
\end{equation}
where we have used $\sum_{i+1}^n (t_s^i)^\alpha/n \to \langle t_s^\alpha\rangle$.
Because the PDF of the residence times on the membrane surface is the same as $\phi(\tau)$, 
 $\langle t_s^\alpha \rangle$ is given by $\langle t_s^\alpha \rangle = \tau_z^\alpha \Gamma (1+\alpha)$. 
Therefore, the effective lateral diffusion coefficient is given by
\begin{equation}
D_{\rm eff} = \frac{D'_s \langle t_s^\alpha\rangle/2 + D_b \mu_b}{\mu_s + \mu_b}.
\end{equation}
Figure~2 shows the MSDs for the exponential and power-law waiting-time PDFs on the surfaces. 
Theoretical results, Eqs. (\ref{msd_exp}) and (\ref{msd_power}), are in good agreement with those of simulations. 
We note that the MSD is always normal because of desorptions. In other words, the MSD does not show a transient subdiffusion 
even when the MSD on the surface is subdiffusive. 
 This is because the initial condition for $z$ direction is in equilibrium.  
Otherwise, the MSD asymptotically exhibits normal diffusion (transient subdiffusion).

\begin{figure}
\includegraphics[width=.9\linewidth, angle=0]{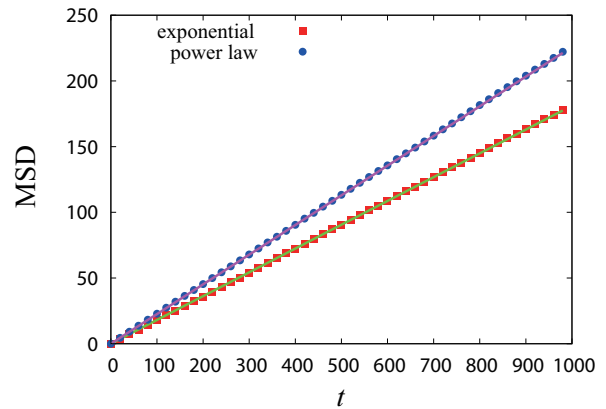}
\caption{Mean square displacement for different waiting time distributions on the membrane surfaces ($\tau_b=1$, $\tau_z=100$, and $L =10$). 
Squares and circles are the results of the numerical simulation in the cases of exponential ($\tau_s=20$) 
and power-law ($\alpha=0.5$ and $\tau_s=1$) waiting-time distributions, respectively. In the power-law waiting-time distribution, we use 
$\psi_s(\tau) = \alpha \tau^{-1-\alpha}$ $(\tau\geq 1)$. 
Solid lines are the theoretical ones.}
\end{figure}

\section{Time-averaged mean square displacement}
Here, we consider fluctuations of the lateral TAMSD on $xy$ plane ($d=2$) in the case of the exponential waiting-time distribution. 
To characterize the fluctuations, we use the RSD, defined by Eq.~(\ref{rsd}). 
For an equilibrium process, the ensemble average of TAMSD coincides with the MSD: 
$\langle \overline{\delta^2(\Delta;t)}\rangle = \langle x_\Delta^2  + y_\Delta^2 \rangle$. 
Therefore, the ensemble average of diffusion coefficients is given by 
\begin{equation}
\langle D_t \rangle 
= \frac{D_s \mu_s + D_b \mu_b}{\mu_s + \mu_b},
\end{equation}
for all $t>0$.

\begin{figure}
\includegraphics[width=.9\linewidth, angle=0]{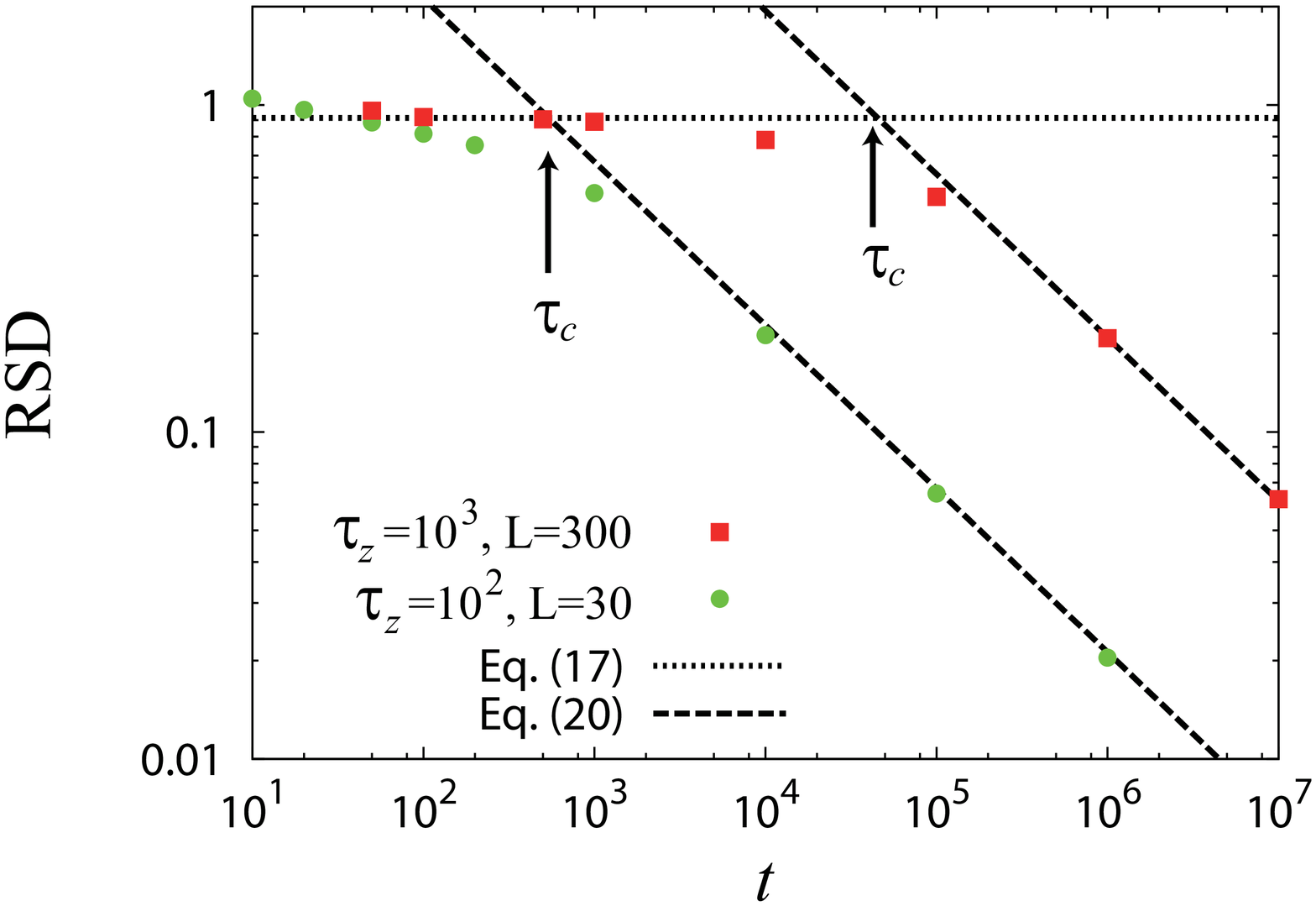}
\caption{Relative standard deviation of the TAMSDs ($\tau_b=1$, $\tau_s=20$, and $L =10$). 
In numerical simulations, we replace $D_t$ as $\overline{\delta^2(\Delta;t)}$ with $\Delta=20$ 
because the TAMSDs show normal diffusion, which is confirmed numerically (not shown). 
The dashed lines are the theoretical ones described by Eq.~(\ref{rsd_short}) and (\ref{rsd_long}),
 where the second moment $\langle \tau_s^2\rangle$ is obtained numerically.}
\label{rsd_crossover}
\end{figure}

As shown in Fig.~\ref{rsd_crossover}, the RSD shows a crossover from a plateau to $t^{-0.5}$ decay. 
This persistent plateau implies a distributional behavior in diffusivity. In other words, an observed diffusivity 
before the crossover remains a random variable.
When the measurement time $t$ is much smaller than the crossover time $\tau_c$, 
the diffusivity is almost determined by that of an initial state, {\it i.e.}, bulk or surface. 
It follows that the probability that diffusion coefficient is $D_s$ is almost equal to $T_s$ and 
the probability that diffusion coefficient is $D_b$ is almost equal to $1-T_s$. Therefore,
for $t\ll \tau_c$, we have
\begin{equation}
\langle D_t^2 \rangle \cong D_s^2 T_s + D_b^2 (1-T_s)= \frac{D^2_s \mu_s + D^2_b \mu_b}{\mu_s + \mu_b}.
\end{equation}
Thus, the RSD is approximately given by
\begin{equation}
\sigma_D(t) \cong \frac{|D_b- D_s|\sqrt{\mu_s\mu_b}}{D_s\mu_s + D_b\mu_b}.
\label{rsd_short}
\end{equation}

Recently, theory of the RSD in a diffusion process with a time-dependent and fluctuating diffusivity has been 
developed \cite{Uneyama2014}. Because the lateral diffusion in our model is a two-state diffusion 
process where the state is randomly fluctuating, we can apply the theory to our model.  
For $t\gg \tau_c$, the theory states that the RSD is given by 
\begin{equation}
\sigma_D^2 (t) \cong \frac{2}{t} \int_0^\infty ds \psi_1(s),
\label{formula}
\end{equation}
where
\begin{equation}
\psi_1(t) \equiv \frac{\langle D_t D_0 \rangle - \langle D_0 \rangle^2}{\langle D_0 \rangle^2}.
\end{equation}
The correlation function in dichotomous processes is calculated in Appendix.~C.
It follows that the RSD for $t\gg \tau_c$ is obtained as
\begin{equation}
\sigma_D(t) \sim \frac{|D_b- D_s|\mu_s\mu_b}{D_s\mu_s + D_b\mu_b} 
\sqrt{\left( \frac{\langle \tau_s^2 \rangle_c}{\mu_s^2} + \frac{\langle \tau_b^2 \rangle_c}{\mu_b^2}\right) 
\frac{1}{(\mu_s + \mu_b)t}},
\label{rsd_long}
\end{equation}
where $\langle \cdot \rangle_c$ is the cumulant. 
Therefore, the crossover time $\tau_c$ from distributional to ergodic behavior is given by 
\begin{equation}
\tau_c = \frac{\mu_s\mu_b}{\mu_s + \mu_b}
\left( \frac{\langle \tau_s^2 \rangle_c}{\mu_s^2} + \frac{\langle \tau_b^2 \rangle_c}{\mu_b^2}\right).
\end{equation}

\section{Method revealing an unknown surface diffusivity}

In experiments, it is difficult to estimate the exact diffusivity on the surface. This is because 
a diffusing particle will desorb from the surface. 
Here, we provide a method revealing the exact diffusivity on the surface, when  the bulk diffusion properties 
are known, {\it i.e.}, the bulk diffusivity $D_b$ and the mean return time $\mu_b$. 
It is important to note that one can obtain the bulk diffusion coefficient $D_b$, the mean return time $\mu_b$ in bulk, 
the effective diffusion coefficient $D_{\rm eff}$, $\sigma_D(0)$, and the crossover time $\tau_c$ by experiments. 
Using $D_{\rm eff}$ and $\sigma_D(0)$, one can know unknown quantities, {\it i.e.}, the surface diffusion coefficient $D_s$ and
the mean trapping time on the surface $\mu_s$. In fact, these quantities are explicitly obtained as
\begin{equation}
D_s = \frac{D_{\rm eff} ( D_{\rm eff} \sigma_D(0)^2 + D_{\rm eff} -D_b)}{D_{\rm eff} -D_b},
\end{equation}
and
\begin{equation}
\mu_s= \frac{\mu_b (D_{\rm eff} - D_b)}{D_{\rm eff}^2 \sigma_D(0)^2}. 
\label{eq:mus}
\end{equation}
In case that $\mu_b$ is unknown, 
we can obtain the ratio $\mu_s/\mu_b$ from Eq. (\ref{eq:mus}). 
Moreover, using $\tau_c$, one can know the cumulant $\langle \tau_s^2 \rangle_c$, {\it i.e.}, the second moment of the trapping time $\tau_s$. 
We note that this crossover time is important to know a characteristic time in the diffusivity. In fact, 
this crossover time is related to the longest relaxation time in entangled polymers \cite{Uneyama2012,Uneyama2014}. 
In general, it is difficult to determine whether a particle is on the surface or not in experiments. Therefore, this is a good method 
to estimate the surface properties because this method does not require a determination of 
whether a particle is on the surface or not. 

\section{Conclusion}
We have shown that the ensemble-averaged MSDs show normal diffusion even when the diffusion on the surface 
is not normal (subdiffusion). Moreover, we find that fluctuations of TAMSDs remain random if the measurement 
time is smaller than a characteristic relaxation time even when the process is in equilibrium. For large measurement times,
the fluctuations  decay as 
$t^{-0.5}$, which is a usual ergodic relaxation. In other words, we find a transition from a non-ergodic (distributional) 
behavior to an ergodic behavior. Although 
a similar phenomenon was found in CTRW where the waiting-time 
distribution has a power law with an exponential cutoff \cite{Miyaguchi2011a}, the transition  in our model results from
 a two-state randomly fluctuating diffusivity, which is a completely different origin from CTRW. 
 Such a two-state diffusion process is an optimal target search process because a searching molecule can bind to 
 the target with the aid of the slow diffusivity near the target (surface). 
We suggest that the transition from non-ergodic 
to ergodic behavior will be universal in optimal target search processes. 

TA thanks T. Uneyama and T. Miyaguchi for discussion about Eq.~(\ref{formula}).
This work was supported in part by Grant for Basic Science Research Projects from The Sumitomo Foundation.

\appendix

\section{Derivation of the PDF of residence times on the membrane surface}

In our model, a random walker desorbs from the membrane surface if the waiting time for the $z$ direction is smaller 
than that for the $xy$ plane. Let $X_i$ and $Y_i$ be random variables with distributions $\psi_s(x)$ and $\phi(y)$, 
respectively. Then, the residence time can be represented by 
\begin{equation}
T= Z_1 + Z_2 + \cdots + Z_{n-1} + Z_n,
\label{RT_surface}
\end{equation}
where $n= \min\{ k \leq 1; Y_k < X_k\}$ and $Z_k$ be the minimum of $X_k$ and $Y_k$: $Z_k \equiv \min (X_k, Y_k)$. 
Using Eq.~(\ref{RT_surface}), we can derive the PDF of the residence times on the membrane surface. 
The PDFs of random variables $Z_k$ with $X_k < Y_k$ and $Y_k < X_k$ are given by
\begin{equation}
\psi_{Z_X}(\tau) = \psi_s(\tau) \int_\tau^\infty \phi(x)dx
\end{equation}
and
\begin{equation}
\psi_{Z_Y}(\tau) = \phi(\tau) \int_\tau^\infty \psi_s(x)dx,
\end{equation}
respectively. The Laplace transform of the PDF $P(T)$ is written as 
\begin{equation}
P^*(s) = \sum_{n=1}^\infty \{ \psi_{Z_X}^*(s)\}^{n-1} \psi_{Z_Y}^*(s)
= \frac{\psi_{Z_Y}^*(s)}{1- \psi_{Z_X}^*(s)}, 
\end{equation}
where $\psi_{Z_Y}^*(s)$ and $\psi_{Z_Y}^*(s)$ are the Laplace transforms of $\psi_{Z_X}(\tau)$ and 
$\psi_{Z_Y}(\tau)$, respectively. Because $\phi(\tau) = \tau_z^{-1} \exp(-\tau/\tau_z)$, the Laplace transforms 
of $\psi_{Z_X}(\tau)$ and $\psi_{Z_Y}(\tau)$ are given by
\begin{equation}
\psi_{Z_X}^*(s) = \psi_s(s+\tau_z^{-1})
\end{equation}
and
\begin{equation}
\psi_{Z_Y}^*(s) = \frac{1}{\tau_z} \frac{1-\psi_s(s+\tau_z^{-1})}{s + \tau_z^{-1}},
\end{equation}
respectively. Therefore, we have 
\begin{equation}
P^*(s) = \frac{1}{1+s\tau_z},
\end{equation}
which means $P(T) = \phi(T)$. 

\section{Lateral subdiffusion in CTRW}

It is known that the MSD in CTRW with the waiting time PDF $\psi_s(\tau) \sim \frac{\alpha}{\tau_s} \left(\frac{\tau}{\tau_s}\right)^\alpha$
($\tau\to\infty$) is given by
\begin{equation}
\langle x_t^2 \rangle \sim \frac{1}{\Gamma(1-\alpha)\Gamma(1+\alpha)} \left(\frac{t}{\tau_s}\right)^\alpha
\end{equation}
for $t\to \infty$ \cite{metzler00}. In our model, if a random walker does not desorb from the surface, a waiting time is assigned 
and it will jump to the $x$ or $y$ direction with equal probability. 
Because the waiting time PDF for the $x$ direction in our model can be written as its convolution, 
the Laplace transform of the PDF is given by
\begin{equation}
\psi^*_x(s) = \sum_{n=1}^\infty \{\psi^*_s(s)\}^n \frac{1}{2^n}
=\frac{\psi^*_s(s)/2}{1-\psi^*_s(s)/2}.
\end{equation}
Using the Laplace transform $\psi_s^*(s)$, we obtain
\begin{equation}
\psi_x^*(s) = 1 - 2\tau_s^\alpha s^\alpha + o(s^\alpha).
\end{equation}
The waiting time PDF for $x$ direction is given by
\begin{equation}
\psi_x(\tau) \sim \frac{\alpha}{2^{\frac{1}{\alpha}}\tau_s} \left(\frac{\tau}{2^{\frac{1}{\alpha}}\tau_s}\right)^{-1-\alpha} 
\end{equation}
for $\tau\to\infty$. 
It follows that the lateral MSD is given by
\begin{equation}
\langle x_t^2 + y_t^2 \rangle \sim \frac{1}{\Gamma(1-\alpha)\Gamma(1+\alpha)} \left(\frac{t}{\tau_s}\right)^\alpha {\color{blue} .}
\end{equation}

\section{Correlation function in dichotomous processes}
We calculate the correlation function, defined by $C(t) \equiv \langle D_t D_0 \rangle - \langle D_0 \rangle^2$, 
in two-state process (dichotomous process). The correlation function is represented by 
\begin{eqnarray}
C(t) &=& D_b^2 P_b W_{bb}(t) + D_bD_s P_s W_{bs}(t)\nonumber\\
&&+ D_sD_b P_b W_{sb} + D_s^2 P_s W_{ss}(t) - \langle D_0 \rangle^2,
\end{eqnarray}
where $P_b$ and $P_s$ are the probabilities of finding a particle initially in the bulk and on the surface, respectively.  
$W_{bb}(t) = \Pr \{D_t=D_b| D_0 =D_b\}$, $W_{bs}(t) = \Pr \{D_t=D_s| D_0 =D_b\}$, $W_{sb}(t) = \Pr \{D_t=D_b| D_0 =D_s\}$,
and $W_{ss}(t) = \Pr \{D_t=D_s| D_0 =D_s\}$ are conditional probabilities, and $\langle D_0 \rangle$ is given by
\begin{equation}
\langle D_0 \rangle = \frac{D_b\mu_b +D_s \mu_s}{\mu_b + \mu_s}.
\end{equation}
The conditional probability can be obtained as
\begin{eqnarray}
W_{bb}(t) &=& \sum_{n=0}^\infty \Pr (N_t=2n) \\
&=& \sum_{n=1}^\infty \{\Pr(S_{2n} < t) - \Pr (S_{2n+1}<t)\}\nonumber\\
&&+ \Pr(N_t=0),
\end{eqnarray}
where $S_n$ is a sum of waiting times, {\it i.e.}, $S_{2n} = {\tau_s^1} + {\tau_b^1} + \cdots + {\tau_s^n} + {\tau_b^n}$. 
Therefore, the Laplace transform is given by
\begin{widetext}
\begin{eqnarray}
\hat{W}_{bb}(s)&=& \frac{1}{s} \sum_{n=1}^\infty \hat{f}_{E,b}(s) \{\hat{\psi}_s(s)\hat{\psi}_b(s)\}^{n-1} \hat{\psi}_s(s)
- \frac{1}{s} \sum_{n=1}^\infty \hat{f}_{E,b}(s) \{\hat{\psi}_s(s)\hat{\psi}_b(s)\}^{n}
+\frac{1-\hat{f}_{E,b}(s)}{s}\\
&=&\frac{1}{s} - \frac{\hat{f}_{E,b}(s)}{s} \frac{1-\hat{\psi}_s(s)}{1-\hat{\psi}_b(s) \hat{\psi}_s(s)},
\label{Wbb}
\end{eqnarray}
\end{widetext}
where $\hat{f}_{E,b}(s)$ is the Laplace transform of the PDF of the forward recurrence time $\hat{f}_{E,b}(\tau)$ when the random walker 
is in bulk, 
which is given by  $\hat{f}_{E,b}(s)=[1-\hat{\psi}_b(s)]/(\mu_bs)$ 
\cite{Cox,Godreche-Luck-2001}. 
In the same way as in Eq.~(\ref{Wbb}), we have
\begin{equation}
\hat{W}_{ss}(s)=\frac{1}{s} - \frac{\hat{f}_{E,s}(s)}{s} \frac{1-\hat{\psi}_b(s)}{1-\hat{\psi}_b(s)\hat{\psi}_s(s)},
\end{equation}
\begin{equation}
\hat{W}_{bs}(s)=\frac{1}{s} - \frac{\hat{f}_{E,b}(s)}{s} \frac{1-\hat{\psi}_s(s)}{1-\hat{\psi}_b(s)\hat{\psi}_s(s)},
\end{equation}
and
\begin{equation}
\hat{W}_{sb}(s)=\frac{1}{s} - \frac{\hat{f}_{E,s}(s)}{s} \frac{1-\hat{\psi}_b(s)}{1-\hat{\psi}_b(s)\hat{\psi}_s(s)},
\end{equation}
where $\hat{f}_{E,s}(s)$ is the Laplace transform of the PDF of the forward recurrence time $\hat{f}_{E,s}(\tau)$ when the 
random walker is on the surface, 
which is given by  $\hat{f}_{E,s}(s)=[1-\hat{\psi}_s(s)]/(\mu_ss)$ 
\cite{Cox,Godreche-Luck-2001}. 
It follows that the Laplace transform of $C(t)$ is given by
\begin{eqnarray}
\hat{C}(s) &=& \left(\frac{D_b - D_s }{\mu_b + \mu_s}\right)^2 \frac{\mu_b\mu_s}{s}\nonumber\\
&&- \frac{(D_b - D_s)^2}{\mu_b + \mu_s} \frac{\{1-\hat{\psi}_b(s)\} \{1-\hat{\psi}_s(s)\}}{\{1-\hat{\psi}_b(s)\hat{\psi}_s(s)\}s^2}.
\end{eqnarray}
Because we assume that all moments of the waiting times are finite,  the Laplace transform becomes 
\begin{equation}
\hat{C}(s) = \frac{(D_b - D_s)^2 (\mu_b\mu_s)^2}{2(\mu_b + \mu_s)^3} \left(\frac{\langle \tau_b^2 \rangle_c}{\mu_b^2} + \frac{\langle \tau_s^2\rangle_c}{\mu_s^2}\right) 
+O(s).
\end{equation}
in the small $s \ll 1$. The integration in Eq.~(\ref{formula}) can be  performed using 
$\int_0^\infty ds \psi_1(s) = \hat{\psi}_1(0) = \frac{\hat{C}(0)}{\langle D_0\rangle^2}$. 
Therefore, we obtain Eq.~(\ref{rsd_long}).

\end{document}